\documentclass[sn-mathphys,Numbered]{sn-jnl}

\usepackage{graphicx}%
\usepackage{multirow}%
\usepackage{amsmath,amssymb,amsfonts}%
\usepackage{amsthm}%
\usepackage{mathrsfs}%
\usepackage[title]{appendix}%
\usepackage{xcolor}%
\usepackage{textcomp}%
\usepackage{manyfoot}%
\usepackage{booktabs}%
\usepackage{algorithm}%
\usepackage{algorithmicx}%
\usepackage{algpseudocode}%
\usepackage{listings}%

\theoremstyle{thmstyleone}%
\theoremstyle{thmstyletwo}%

\theoremstyle{thmstylethree}%

\begin{document}

\title[Fukushima tritiated water release]{Fukushima tritiated water release\\-- What is the polemic all about?}


\author{\fnm{Hans~}Peter \sur{Beck}}

\affil{\orgdiv{Laboratory of High Energy Physics}, \orgname{University of Bern}, \orgaddress{\street{Sidlerstr.~5}, \city{Bern}, \postcode{3012}, \country{Switzerland}}}

\date{\today}



\abstract{
A mere amount of $2.2~\mathrm{grams}$ ($780\mathrm{~TBq}$) of tritium, diluted in  $1.25 \cdot 10^6~\mathrm{m^3}$ water, contained in 1047 tanks at the Fukushima Daiichi 
nuclear power plant are being released to the Pacific Ocean. The operation is scheduled to last over $30~\mathrm{years}$, with not more than releasing $62~\mathrm{mg}$ ($22~\mathrm{TBq}$) 
of tritium annually.
The outcry in the world’s press and the world’s population is  huge and countries like e.g. China are protesting aloud and are even banning Japanese seafood being sold in their 
domestic market. The outcry is real, the perceived fears are real, the havoc created on the Japanese fish market is real, but the danger is non-existing. The panic results from 
over-regulations initiated by the International Commission on Radiological Protection (ICRP) and similar bodies worldwide, prohibiting a reliable assessment of dangers and are 
thereby also preventing a solid risk analysis of real dangers.
}




\maketitle

\section{The tsunami that changed it all}\label{sec1}

On 11 March 2011 a magnitude 9.1 undersea earthquake in the Pacific Ocean triggered a tsunami that with a height of 14~m when reaching Japan’s Pacific coastline brought unbound destruction. 
20’000 lives were lost, entire towns were devastated, and the tsunami was also at the cause of the Fukushima Daiichi nuclear disaster \cite{JpGov1}. The nuclear reactors shut down automatically upon 
registration of the earthquake, but the reactor cores still needed ongoing cooling. Flooding of the area caused the failure of the emergency power generators and resulted in a loss of reactor 
core cooling that finally led to nuclear meltdowns. The released heat from the meltdown was at the cause of hydrogen explosions, where reactor core material from three reactor cores was 
carried into the atmosphere or directly washed out into the ocean. \\

Regarding radiation exposure, 96\% of the workers at Fukushima Daiichi NPP were exposed to less than 50~mSv. A total radiation dose of greater than 200~mSv was observed in nine workers. 
Of these, two workers were exposed to greater than 600~mSv, with 679~mSv being the highest. There were no deaths from radiation exposure in the immediate aftermath of the incident \cite{Hasegawa2016}.

\section{The problems with Linear No-Threshold and ALARA}

Over 110’000 residents in the surrounding areas were evacuated, causing 2’268 non-radiation disaster-related deaths due to many stress factors implied~\cite{JpGov2}. On the long-term impact, the 
maximum predicted eventual cancer mortality and morbidity estimate according to the scientifically flawed and heavily disputed linear no-threshold (LNT) hypothesis is about 1800 
residents \cite{C2EE22019A}. The LNT hypothesis of ionizing radiation–induced mortality and morbidity assumes that every increment of radiation dose, no matter how small, constitutes an increased 
cancer risk for humans. LNT is presently the most widely applied model for radiation risk assessment. However, no adverse health effects among Fukushima residents 
have been documented that are directly attributable to radiation exposure from the Fukushima Daiichi nuclear power station accident \cite{unscear}. LNT is at the base of all radioprotection measures 
and regulations, but LNT completely ignores the body’s capability to heal any damage made to any of its cells~\cite{RICCI2019128, THARMALINGAM201954, CALABRESE2023110653, OConnor:2017aa}. LNT’s base assumption is 
that radiation damage accumulates over time without any healing process taking place. LNT also turns a blind eye to the rate at which radiation is absorbed, and diligently ignores whether a given amount 
of radiation is absorbed in a fraction of a second or is accumulated evenly over the full course of a year. \\

A wide range of literature exists, which is based on empirical data collected over decades,  shows LNT overestimates effects of low-level radiation by orders of magnitude - 
see e.g.~\cite{RICCI2019128, THARMALINGAM201954, CALABRESE2023110653, OConnor:2017aa} and references mentioned therein. 
The Swiss Federal Office for  Public Health states in \cite{BAG} still carefully that “\textit{the minimum dose at which an effect can be detected varies according to the observation of the collective and is around 100~mSv}”
and Ref.~\cite{THARMALINGAM201954} reports that low dose rates are even beneficial, in stark contrast to the LNT hypothesis: \textit{"low dose rates improve tumour suppression, inhibit cancer formation and protect 
against neoplastic transformation"}, which is also confirmed in Ref.~\cite{RICCI2019128}.\\

One can argue that LNT’s intention is to be safe, and radiation shall be as low as reasonably achievable, also called ALARA \cite{HANSSON2013143}. However, ALARA ignores known empirical molecular biology 
data~\cite{RICCI2019128, THARMALINGAM201954, OConnor:2017aa} and ignores other risk factors other than radiation exposure that need to be factored in when making decisions to mitigate harm or when defining regulations to mitigate risks. 
One easily comes to the  conclusion that the evacuation radius around the Fukushima Daiichi plant was way too big and that evacuated people could have reoccupied their homes rather quickly within weeks, 
after the most active short-lived nuclides have decayed, and that the current LNT and ALARA based scheme created a lot of harm unnecessarily.

\section{Clean-up work}

\begin{figure}[H]
\centering
\includegraphics[width=1.0\textwidth]{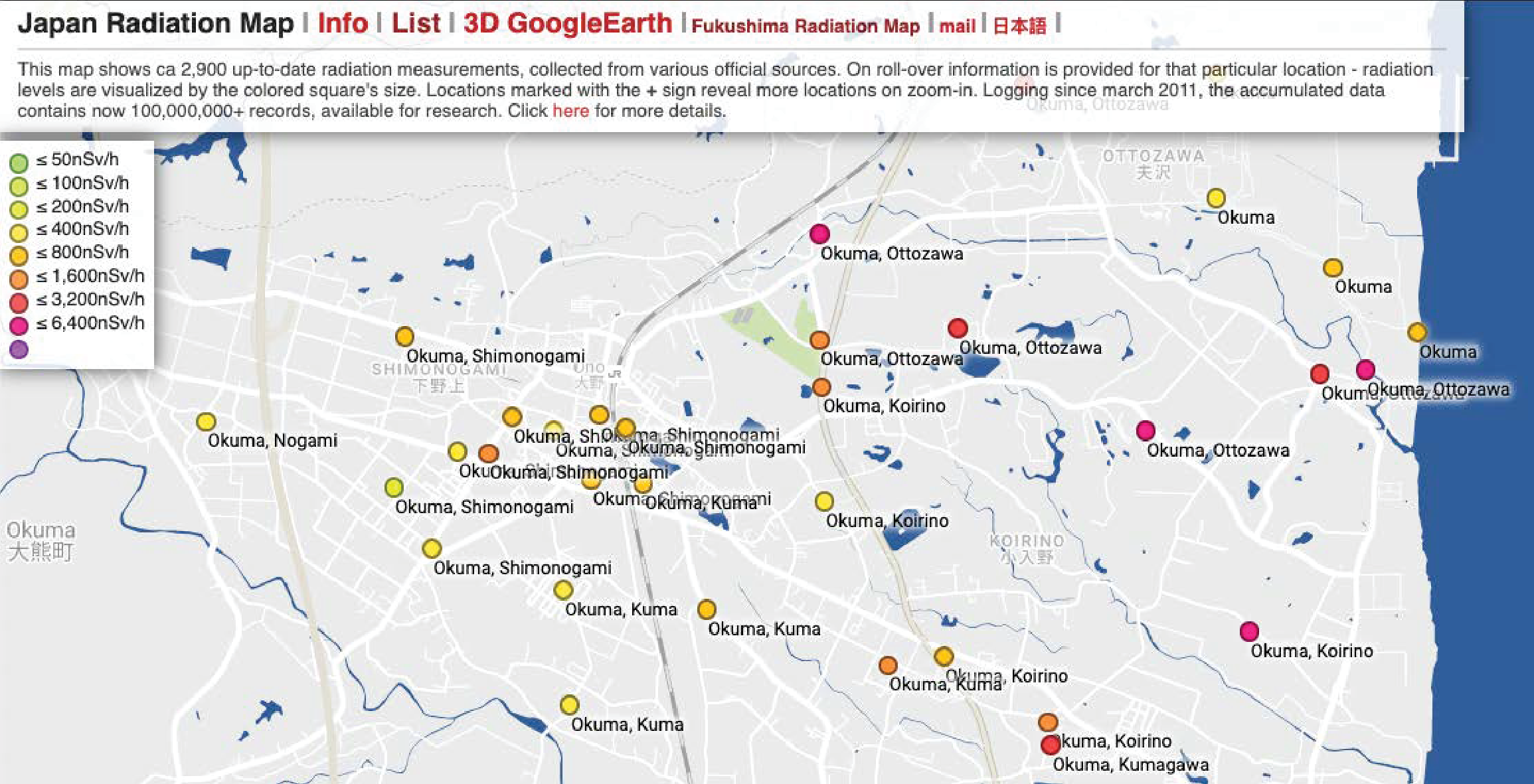}
\caption{Japan radiation map as of 1 September 2023, showing radiation measurements in the Fukushima region \cite{JpRadMap}.}\label{FigRadMap}
\end{figure}

The cleaning of the area around the Daiichi nuclear plant is ongoing still today. Radiation levels have become acceptable in most places or, in some places, are not higher than when flying at cruise 
altitude in a passenger airplane ($2000-7000~\mathrm{nSv/h}$, depending on latitude), Fig.~\ref{FigRadMap}.  \\

Water that was used to cool the melted-down reactor cores got contaminated and is still stored in large tanks at the Daiichi site \cite{TEPCO}. Radionuclides in the contaminated water can be and are 
filtered out, but this is not possible for the tritium dissolved in water. Tritium (T) is chemically identically to hydrogen (H) and binds to oxygen (O), forming tritiated water T-O-H and T-O-T, 
chemically non-distinguishable from ordinary water H-O-H.  \\

A slow release into the ocean of diluted tritiated water with an activity below $1500~\mathrm{Bq/\ell}$ is the agreed way to dispose of the tritiated water, where WHO defines $10’000~\mathrm{Bq/\ell}$ as 
tolerable for drinking water \cite{WHO}. As an alternative solution, vaporization of the tritiated water into the atmosphere was discussed but not pursued. TEPCO, the Tokyo Electric Power Company that operates the 
Daiichi nuclear plant and responsible for the cleaning and decontamination, plans to cap the annual level of tritium released at 22 trillion becquerels [$22\cdot10^{12}~\mathrm{Bq}$ or
$22~\mathrm{TBq}$] annually over more than 30 years \cite{TEPCO}. \\

\begin{figure}[H]
\centering
\includegraphics[width=0.75\textwidth]{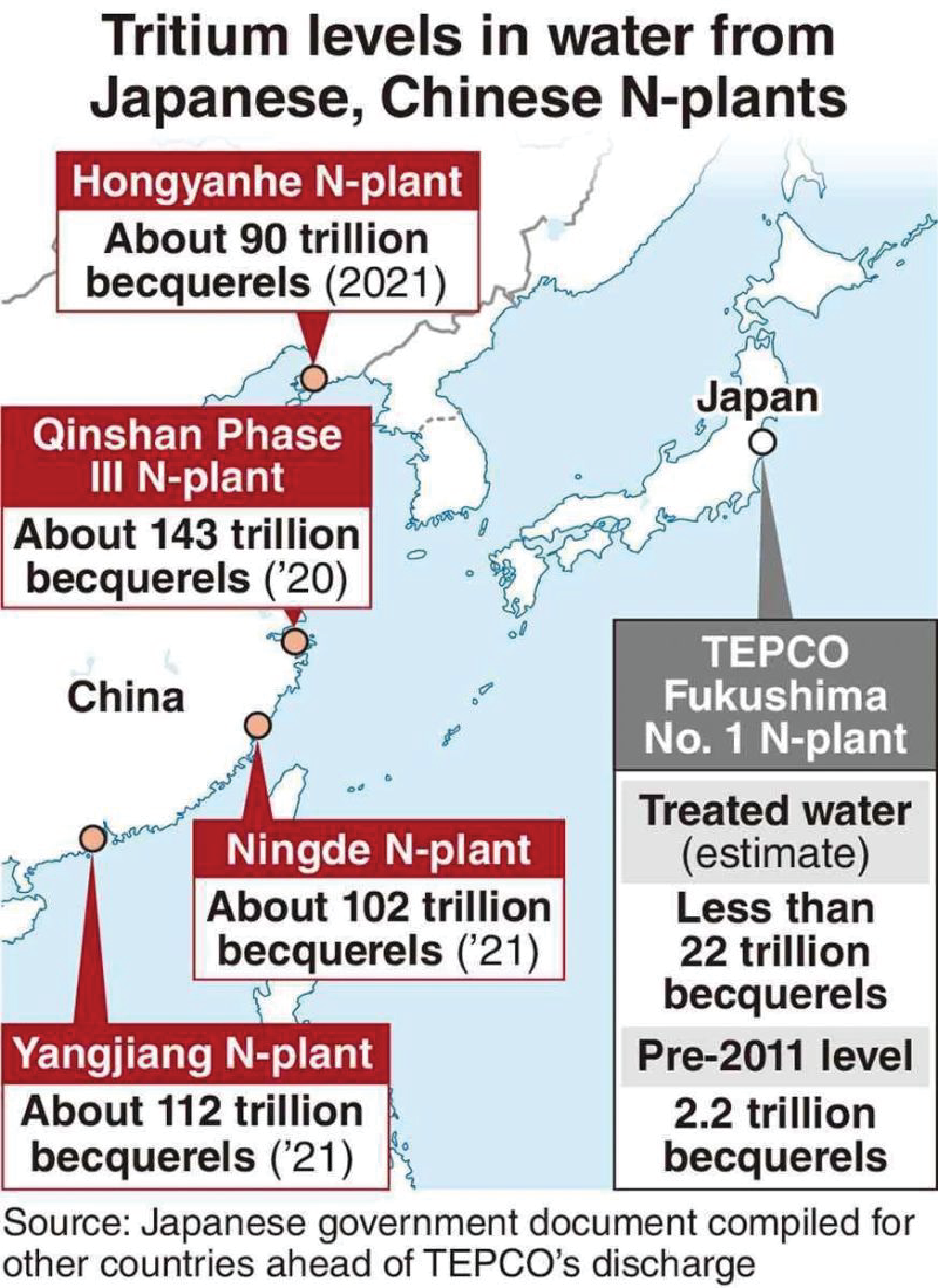}
\caption{Annual release of tritiated water in China and at the Fukushima Daiichi site in Japan \cite{TritRel}.}\label{FigTritRel}
\end{figure}

This release, which started on 24 August 2023, is heavily discussed, and was leading to headlines in major newspapers worldwide. However, releasing tritiated water is common practice, 
where China who openly condemns Japan for its release of tritiated water is not so quite innocent and regularly releases tritiated water at an order of magnitude 
higher than the Fukushima tritiated water release, as is reported by the Japanese Government and illustrated in~Fig.~\ref{FigTritRel}.\\

\section{Estimating the dangers}

The question arising is, what is exactly the danger related to tritiated water with an activity of $1500~\mathrm{Bq/\ell}$ and of releasing $22~\mathrm{TBq}$ annually to the ocean? \\

Tritium is a beta emitter with a half-life of 12.32~years. The beta particle, i.e. the decay electron, has an endpoint energy of 18.6~keV, with an average energy of 5.7~keV. Such electrons 
are immediately slowed down in water and even at their maximum endpoint energy, their maximum range in water is less than 7~$\mu$m before they come to a standstill and can’t do 
any harm anymore. Swimming in tritiated water is completely harmless, as decay electrons are not capable of penetrating the skin's epidermis, which has a thickness of $30-50~{\mu}\mathrm{m}$, 
even then when a tritium decay shall happen in the ultimate direct vicinity and targeted at $90^\circ$ to the swimmer’s bare skin. Nor can such decay electrons penetrate through fish scales. \\

Drinking tritiated water is, however, different as the decay electron is ultimately absorbed inside the body. When ingesting tritiated water, the body takes it as ordinary water and passes 
it through the body’s digestive system, where some of it passes fast through the metabolic system and some of it is deposited inside the body’s cells, and  released after 
some time. A biological half-time of 10~days for 99.00\%, of 40~days for 0.98\% and of 350~days for 0.02\% of of the ingested tritium is reported in Ref.~\cite{Bundy2012}.
A biological half-time of tritium in a human body of  $\mathrm{T}_{1/2}^\mathrm{bio} = 12$~days is considered here for simplicity, which covers by a good margin the absorbed dose rate of 
the small fraction of the ingested tritium that stays long inside the human body. \\

Drinking one litre of tritiated water with $1500~\mathrm{Bq/\ell}$ equates to 
\[
N_0=\frac{A}{\lambda}
   =\frac{A}{\frac{\ln 2}{\mathrm{T}_{1/2}}}
   =\frac{1500~\mathrm{s^{-1}}}{\ln 2} \cdot 12.32 \cdot 365.25 \cdot 86400~\mathrm{s} 
   = 841\cdot 10^{9}
   \]
ingested tritium nuclei. Where $A=1500~\mathrm{s^{-1}}$ is the activity,  $\lambda$ the decay constant, and $\mathrm{T}_{1/2}=12.32~\text{years}$ is the half-life  of tritium that needs 
to be converted in units of second. \\

The number of tritium decays, during the time these ingested nuclei reside in the body, follows as
\[
N_\mathrm{decay}= \int_0^\infty \lambda \; N_0 \;e^{-\lambda_\mathrm{bio}t} \; e^{-\lambda t} \; dt
                = N_0 \,  \frac{\lambda}{\lambda_\mathrm{bio} + \lambda}           
                = 0.0027 \cdot N_0 
                = 2.2 \cdot 10^{9},
\]
where $\lambda_\mathrm{bio}=\frac{\ln 2}{\mathrm{T}_{1/2}^\mathrm{bio}}=6.7\cdot 10^{-7}~\mathrm{s^{-1}}$ is the biological decay constant, defining the rate at which tritium 
is washed out from the human body. With an average energy of $E_\mathrm{avg}=5.7~\mathrm{keV}$ per tritium decay, a total energy dose of  
$E
=
N_\mathrm{decay} \cdot E_\mathrm{avg}
=
12.8\cdot 10^{12}~\mathrm{eV}
=
2.0\cdot 10^{-6}~\mathrm{J}$ is thereby absorbed. \\

Assuming a person to be 80~kg of mass, and the tritium evenly distributed, this results in an energy dose of 
$D=E/80~\mathrm{kg}=26\cdot 10^{-9}~\mathrm{J\;kg^{-1}} = 26 \cdot 10^{-9}~\mathrm{Gray}$. 
The radiative weighting factor for electrons is just one, such that this number is also its equivalent value in units of Sievert. We can therefore conclude that drinking one litre of tritiated
 Fukushima water results in an absorbed dose of 26~nSv.
 
 If the integration time is limited to the first 24~hours only (rather than to infinity, as performed here above), the number of decays during the first day after ingestion follows as 
 $N_\mathrm{decay}^\mathrm{24h}
=
N_\mathrm{decay} \cdot (1-e^{-(\lambda_\mathrm{bio}+\lambda)\cdot 24\mathrm{h}})
=
N_\mathrm{decay}  \cdot 5.6\% 
=
0.13\cdot10^{9}$
 , leading to an exposure of 1.44~nSv in the first day, and correspondingly, to 1.36~nSv in the following day, etc. \\
 
Understanding these values needs some context. For instance, banana are radioactive beta emitters themselves, due to their relatively high level of potassium, and with it, its radioactive variant potassium-40. 
Eating a single banana leads to an exposure of 100~nSv, which is often referred to as the banana equivalent dose \cite{banana}. Eating a single banana, therefore, corresponds 
already to drinking four (!) litres of tritiated Fukushima water. \\

In Switzerland, an average person is exposed to 6.1~mSv annually from environmental radiation and medical diagnostic, corresponding to an average rate of 750~nSv/h \cite{ENSI}. 
This means that drinking 1 litre of tritiated water corresponds to about 120~seconds of an average person’s exposure in Switzerland. In turn, when flying at cruise altitude in 
a commercial airplane, the dose rate is tenfold, due to cosmic radiation. Therefore, drinking one litre of tritiated water corresponds to a few 10s 
of seconds flight at cruise altitude. \\

If we took the non-diluted tritiated water instead, which is $1.25\cdot10^6$~m$^3$ of tritiated water at an average concentration of $620~\mathrm{kBq/\ell}$ and resulting to a total amount of $780~\mathrm{TBq}$,
which by the way corresponds to a mere 2.2 grams (!) of pure tritium, that are being released \cite{TEPCO}, things become slightly different — but are still not alarming. \\

Drinking one litre of non-diluted tritiated water with $620~\mathrm{kBq/\ell}$, results to an exposure of $11~\mu\mathrm{Sv}$, equivalent to 16 hours of average exposure in Switzerland, or a 100 minute flight, 
or eating 11 bananas, which can be spread to eating one banana a day for 11 days. \\

Releasing $22~\mathrm{TBq}$ annually ($62~\mathrm{mg}$ per year (!))  into the ocean, where the tritiated water quickly dilutes to extremely small values does not cause harm in any way. 
Plastic and other toxic chemical waste that finds its way unhindered into the world’s seas in turn are of a real concern. The tritium vanishes with a half-life of 12.32~years, where 
toxic chemicals and other waste stay. \\

The time it takes to release the full amount of the tritiated water can be estimated by $780~\mathrm{TBq} \cdot(1-\alpha t) \cdot e^{-\lambda t} < 22~\mathrm{TBq}$, where
 $\alpha=22/780~\mathrm{year^{-1}}$ is the annual release rate, $t$ is time, and $\lambda$ is the tritium decay rate. 
 When the remaining amount has reached  $22~\mathrm{TBq}$ the final year has come. After $\sim 31$ years, all the tritiated water will be released, four years shorter than a simple estimate from 
 dividing  $780~\mathrm{TBq} \; / \;22~\mathrm{TBq \; year^{-1}}= 35.5~\mathrm{years}$ would result in and taking into account the tritium that will decay while still being inside the tanks.

\section{A brief history of radiation protection}

Drinking the full amount of $2.2~\mathrm{g}$ tritium ($780~\mathrm{TBq}$) leads to a lethal dose.

\[
N_0=\frac{A}{\lambda}
   =\frac{780\cdot 10^{12}~\mathrm{s^{-1}}}{\ln 2} \cdot 12.32 \cdot 365.25 \cdot 86400~\mathrm{s} 
   = 4.4\cdot 10^{23}
\]

 tritium nuclides would be absorbed, of which a small fraction  decays in the first 24 hours:

\begin{equation*}
\begin{split}
N_\mathrm{decay}^\mathrm{24h}
             & = \int_0^\mathrm{24h} \lambda \; N_0 \;e^{-\lambda_\mathrm{bio}t} \; e^{-\lambda t} \; dt
                = N_0 \; \frac{\lambda}{\lambda_\mathrm{bio} + \lambda} \cdot (1-e^{-(\lambda+\lambda_\mathrm{bio})\cdot\mathrm{24h}}) \\
              & = 0.15\cdot10^{-3} \cdot N_0 
                = 6.5 \cdot 10^{19}.
\end{split}
\end{equation*}
          
These lead to an absorbed energy in the first day of

$E^\mathrm{24h}
=
N_\mathrm{decay}^\mathrm{24h} \cdot E_\mathrm{avg}
=
3.7\cdot 10^{23}~\mathrm{eV/day}
=
60~\mathrm{kJ/day}$
, corresponding to an absorbed dose of 
\[
D^\mathrm{24h}=E^\mathrm{24h}/80~\mathrm{kg}=745~\mathrm{J\;kg^{-1}\;day^{-1}} = 745~\mathrm{Gray/day} = 745~\mathrm{Sv/day}.
\]

This dose is definitely and undisputedly deadly! - But what dose would be tolerable not causing harm? \\

Defining what levels of absorbed dose are acceptable and non-harming has a long history, of which here only the time after the end of World War II is briefly recapitulated. 
In 1951, the International Commission on Radiological Protection (ICRP) dose rate limit for the general public was set to 4.4 mSv/week (which was defined as 0.5 Röntgen 
per week in the then used units), which leads to an average of \mbox{0.63~mSv/day}~\cite{ICRPhist}.
From this, the then tolerated dose would have resulted in ingesting 1.8~$\mu$g tritium ($0.66~\mathrm{GBq}$) to be safe. This amount of tritium is contained in 1060 litres of 
concentrated tritiated water inside the Fukushima tanks - or in 440’000 litres of the diluted tritiated water at  $1500~\mathrm{Bq/\ell}$ — still at the tolerable radiation dose 
as they were valid in 1951 ! \\

Ever since 1951, ICRP’s recommendations were made more stringent - never on empirical data, but always based on the LNT hypothesis and the ALARA principle.
Indeed, in the report of Sub-Committee I in the 1954 recommendations, it was stated that ‘\textit{since no radiation level higher than the natural background can be regarded 
as absolutely ‘‘safe’’, the problem is to choose a practical level that, in the light of present knowledge, involves a negligible risk}’. However, the Commission had not rejected 
the possibility of a threshold for stochastic effects \cite{ICRPhist}.

In 1959, ICRP’s publication 1 \cite{ICRPpub1} appeared,  defining a new limit of $50~\mathrm{mSv/y}$ for nuclear workers and $5~\mathrm{mSv/y}$  for the public.
ICRP’s publication 9 \cite{ICRPpub9}, recommended in 1966 that ‘\textit{all doses be kept as low as is readily achievable, economic and social consequences being taken into 
account}’ and publication 22 \cite{ICRPpub22} reported in 1973 that '\textit{the optimum level of protection might be found by means of differential cost–benefit analysis and that the principle
described in Paragraph 52 of Publication 9 was the principle of optimisation of protection.}', which is referring to  \textit{keep radiation doses as low as is readily achievable}. Publication 26 \cite{ICRPpub26} 
from 1977 finally set the limit to  $5~\mathrm{mSv/y}$ for nuclear workers and $1~\mathrm{mSv/y}$ for the public, again based on principles of reducing radiation as much as 
possible, arguing on ethical  grounds, under the  assumption of the LNT hypothesis. \\
 
 Applying $1~\mathrm{mSv/y}$  results in an average of $2.7~\mu\textrm{Sv/d}$. Hence, ingesting  8~ng  of tritium is still today considered tolerable, if no other exposures are assumed. 
 This allows drinking 4.6~litres of the concentrated tritiated water at $620~\mathrm{kBq/\ell}$, or drinking of close to 2’000 litres of the diluted tritiated water at $1500~\mathrm{Bq/\ell}$ ! \\
 
 Certainly, keeping radiation doses low is good intention of ICRP and similar bodies. However, if the derived limits are not based on scientifically, empirically 
 collected data, but are based only on assumptions such as LNT and on a principled fear, regulations become biased towards other avoidable hazards. 
 The non-necessary evacuation of 110’000 people following the Fukushima accident and causing stress-related deaths, where zero radiation-related deaths are to be 
 mourned, is such an example~\cite{JpGov2}. 
 
\begin{figure}[h]
\centering
\includegraphics[width=1.0\textwidth]{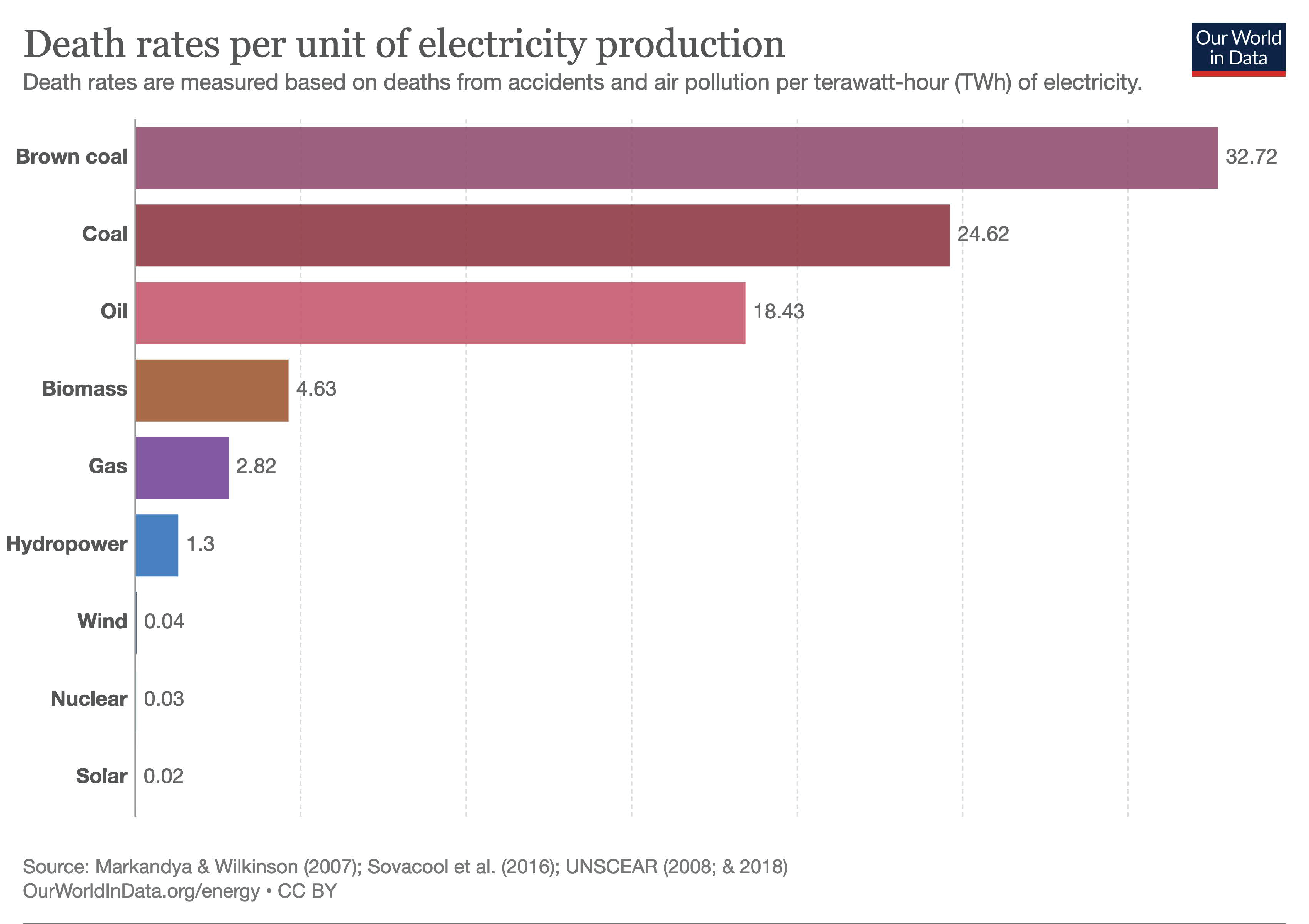}
\caption{Mortality from electricity production by energy source \cite{owidenergy}.}\label{FigMortality}
\end{figure}

 Another example follows from the turning off of functional electricity-producing nuclear power plants from a principled fear of radiation, as e.g. Germany does. To compensate for the 
 lost electricity production, Germany reactivates old coal plants~\cite{bloomberg}, emitting huge amounts of  $\mathrm{CO_2}$ into the atmosphere. Although filtering systems are employed at modern coal plants,
the environment is still polluted with fly ashes that are not only also radioactive themselves, but due to micro particles  released, are at the cause of lung diseases, cardio-vascular problems  and premature deaths, as is shown 
in Fig.~\ref{FigMortality} and discussed  in~\cite{owidenergy}.  Over 24 (30) premature deaths are to be mourned  per TWh of generated electricity from coal (brown coal), but this could be $\mathcal{O}(1000)$ times less if electricity is produced 
from nuclear power instead, where disasters like Chernobyl and  Fukushima have already been accounted for~\cite{owidenergy}. \\

The LNT hypothesis and the ALARA principle deserve a deep reevaluation, and possibly need to be abolished to make place to new and better suiting 
regulations~\cite{RICCI2019128, THARMALINGAM201954, CALABRESE2023110653, OConnor:2017aa}. 
Indeed, no radiation harming effects can be detected below \hbox{$\sim100~\mathrm{mSv}$~\cite{BAG}}, and Refs.~\cite{RICCI2019128, THARMALINGAM201954}, to name only two,
report about beneficial effects of low-dose-rate radiation.
ICRP’s publication 9 \cite{ICRPpub9}, recommended in 1966 that ‘\textit{economic and social consequences being taken into account}’, when it introduced the ALARA principle.
It seems that \textit{economic and social consequences} have been forgotten all about when evacuations are ordered unnecessarily, when nuclear power plants have to give way 
to reactivating old coal plants, or when the resources spent to handle 2.2 grams of tritium are nowhere in balance to the extremely limited danger these 2.2 grams effectively  pose.
 
 \section{Conclusion}
 
 In conclusion, the Fukushima disaster could have been prevented with better protection of the emergency cooling system. The core meltdowns caused wider damage that, 
 however, is still a local disaster, but not a world crisis \cite{C2EE22019A, unscear}, where Ref.~\cite{C2EE22019A} calculates mortality 
 and morbidity numbers for various world regions based on LNT, while also criticising its use at low-dose rates. The evacuation of the surrounding population was causing severe 
 stress-related harm~\cite{Hasegawa2016} and  was most likely unnecessary.  This is easy to say in the aftermath, but  a different and more reasonable approach 
 could have been taken if LNT, ALARA, and the principled fear of radiation would not have prevented proper risk assessment and sensible decision taking.\\

The release of $2.2~\mathrm{g}$ ($780~\mathrm{TBq}$) of  tritiated water is of zero concern at the current rate and activities - $62~\mathrm{mg/a}$ ($22~\mathrm{TBq/a}$) over more than 30 years. 
Releasing tritiated water is also a standard procedure done by many countries, including China. The released water is of drinking water quality, as defined by WHO~\cite{WHO}.\\

Tritium is produced naturally when cosmic rays penetrate through the atmosphere, and therefore, water collected from rain has a tiny activity of  about $1-2~\mathrm{Bq/\ell}$ just from tritium alone~\cite{Palcsu:2018aa}. 
These, together with other radiocative nuclides constantly produced in the high atmosphere, enter the oceans when it rains, these also rain on our heads and these we also drink 
on a daily basis. The annual global precipitation volume is about $5 \cdot 10^{14}~\mathrm{m^3}$~\cite{precipitation}, and hence, $\mathcal{O}(10^6)~\mathrm{TBq}$ of tritium rains on Earth annually, dwarfing the 
Fukushima release by five orders of magnitude. \\

Plastic and other toxic chemical waste that finds its way unhindered into the world’s seas are in turn of a real concern. The tritium vanishes with a half-life of 12.32 years, where 
toxic chemicals and other industrial waste stays. \\

Over-regulations from the strict adoption of the flawed linear no-threshold hypothesis and the resulting ALARA principle are at the cause of creating fears that are unnecessary. 
These are also prohibiting proper risk analyses and the scientific assessment of real dangers. Proper risk assessments are vital in times when decisions need to be taken.
An urgent revision of the radio-protection regulatory framework is not easy to achieve, but given the havoc it produces worldwide, it is of a pure necessity.

\bibliography{Fukushima_tritiated_water_release}

\end{document}